# Scalable Etch-Free Transfer of Low-Dimensional Materials from Metal Films to Diverse Substrates

*Kentaro Yumigeta [a†], Muhammed Yusufoglu [a†], Mamun Sarker [b†], Rishi Raj [c], Franco Daluisio [d], Richard Holloway [a], Howard Yawit [a], Thomas Sweepe [a], Julian Battaglia [a], Shelby Janssen [a], Alex C. Welch [d], Paul DiPasquale [a], K. Andre Mkhoyan [c], Alexander Sinitskii [b], Zafer Mutlu [a,e,f,]\**

[a] *Department of Materials Science & Engineering, University of Arizona, Tucson, Arizona 85721, USA*

[b] *Department of Chemistry, University of Nebraska-Lincoln, Lincoln, Nebraska 68588, USA*

[c] *Department of Chemical Engineering and Materials Science, University of Minnesota, Minneapolis, MN 55455, USA*

[d] *Department of Chemical and Environmental Engineering, University of Arizona, Tucson, Arizona 85721, USA*

[e] *Department of Electrical and Computer Engineering, University of Arizona, Tucson, Arizona 85721, USA*

[f] *Department of Physics, University of Arizona, Tucson, Arizona 85721, USA*

*† These authors contributed equally to this work.*

*\* Corresponding Author: zmutlu@arizona.edu*

## Abstract

Low-dimensional materials hold great promises for exploring emergent physical phenomena, nanoelectronics, and quantum technologies. Their synthesis often depends on catalytic metal films, from which the synthesized materials must be transferred to insulating substrates to enable device functionality and minimize interfacial interactions during quantum investigations. Conventional transfer methods, such as chemical etching or electrochemical delamination, degrade material quality, limit scalability, or prove incompatible with complex device architectures. Here, a scalable, etch-free transfer technique is presented, employing Field's metal (51% In, 32.5% Bi, and 16.5% Sn by weight) as a low-melting-point mechanical support to gently delaminate low-dimensional materials from metal films without causing damage. Anchoring the metal film during separation prevents tearing and preserves material integrity. As a proof of concept, atomically precise graphene nanoribbons (GNRs) are transferred from Au(111)/mica to dielectric substrates, including silicon dioxide ($SiO_2$) and single-crystalline lanthanum oxychloride (LaOCl).

Comprehensive characterization confirms the preservation of structural and chemical integrity throughout the transfer process. Wafer-scale compatibility and device integration are demonstrated by fabricating GNR-based field-effect transistors (GNRFETs) that exhibit room-temperature switching with on/off current ratios exceeding $10^3$. This method provides a scalable and versatile platform for integrating low-dimensional materials into advanced low-dimensional materials-based technologies.

## Introduction

Low-dimensional materials exhibit exceptional electronic and optical properties, making them essential for overcoming the limitations of silicon-based devices and enabling the next generation of advanced technologies [1–4]. However, their synthesis often requires specific substrates and high temperatures, causing significant challenges for practical applications [5]. Transfer methodologies that maintain structural and functional integrity of these materials during processing offer a solution to bridge the gap between synthesis conditions and device integration requirements [6].

Among low-dimensional materials, bottom-up synthesized carbon nanostructures, especially graphene nanoribbons (GNRs), offer distinct advantages due to their atomically precise structural control and tunable properties [7–20]. Their synthesis typically relies on catalytic reactions on metal surfaces (e.g., Au, Cu, and Ag), making it essential to transfer them to insulating substrates for device integration. In particular, Au surfaces are commonly used as growth substrates for carbon nanostructures as well as for various other low-dimensional materials, including transition metal dichalcogenides (TMDs) [21–28], hexagonal boron nitride (hBN) [29], iron oxide (FeO) [30], metal halides [31–34], metal-organic frameworks (MOFs) [35–37], and polymers [38]. Metal thin films, including Au, are attractive as synthesis platforms due to their scalability and cost-effectiveness, and ability to produce single-crystal metal layers with well-defined crystallographic orientations when grown epitaxially on crystalline substrates, enabling the production of high-quality, large-area films of functional materials without grain boundaries [22,39–41].

Despite these advantages, existing transfer methods struggle to maintain material quality at scale. Two main approaches are currently employed for transferring materials grown on metal surfaces: chemical approaches (e.g., metal etching) and physical approaches (e.g., electrochemical delamination). Metal etching approach involves the use of aggressive etchants to dissolve metal thin films, enabling large-area processing but often causing irreversible chemical damage that degrades material quality [42,43]. This approach leads to significant risks, including unwanted modification of electronic properties in sensitive nanomaterials such as magnetic GNRs, unintentional doping, and incompatibility with established device fabrication processes. On the other hand, electrochemical delamination avoids corrosive chemicals by using mechanical stress generated during electrochemical hydrogen evolution [44–46]. However, this approach requires high mechanical stability of a metal thin film and its strong adhesion to an underlying substrate so that the metal is not delaminated together with the targeted functional material by hydrogen bubbles. These requirements are not met by typical metal films, such as thin Au films on mica that are commonly used for the on-surface growth of carbon nanostructures. Although metal thin films provide excellent catalytic surfaces for the growth of atomically precise structures, their weak adhesion to substrates and mechanical fragility limit the substrate materials, as well as thickness, and size, making it unfeasible for wafer-scale transfer.

To address these limitations, previous research efforts on improving transfer processes have primarily focused on conventional materials such as polymers and commonly used metals [6]. In this study, we present a wafer-scale, etch-free transfer method using low-melting-point metals (LMPMs) as a mechanical support for metal thin films during electrochemical delamination. LMPMs provide strong adhesion to metal thin films, enabling wafer-scale transfer without wet etchants. This approach provides several advantages: reusability of Au films for subsequent material synthesis, reduced environmental impact through minimized chemical waste, low-temperature processing (below 100°C) to avoid thermal damage to sensitive materials, and recyclability through repeated melting and solidification cycles, making it practical for industrial applications. To demonstrate this methodology, we selected Field's metal as the LMPM. As a proof of concept, atomically precise GNRs, which require an Au surface for their growth, were transferred via the electrochemical delamination process, confirming the structural uniformity and minimal defect density of the transferred GNRs. Full-wafer transfer capability was validated through the delamination of polymer films from 100-mm diameter Au thin films.

## Results and Discussion

For this demonstration of the LMPM-assisted transfer procedure, we selected Field's metal (51% In, 32.5% Bi, 16.5% Sn by weight) due to its low melting point of 62°C and its cadmium- and lead-free composition, which offers reduced toxicity compared to other low-melting-point alloys. The basic procedure for the transfer process is illustrated in Figure 1.

First, an Au thin film is deposited onto a substrate, and low-dimensional materials are grown on top of it (Figure 1a). A polymer film is then spin-coated onto the low-dimensional material and Au thin film before detaching them from the substrate (Figure 1b and c). Liquid LMPM is applied to the Au film side and solidified (Figure 1d). The high conductivity and strong adhesion of LMPM prevent damage to the Au thin film during hydrogen bubble delamination at the polymer/metal interface (Figure 1e). As a result, carbon nanostructures along with the polymer layer are successfully separated using this method and transferred onto a target substrate (Figure 1f); finally, the polymer is removed by solvent (Figure 1g).

After delamination, the Au thin film remains on the LMPM, and the majority of the Au film can be recovered from the molten LMPM. Furthermore, LMPMs can be melted and solidified repeatedly for reuse in multiple transfer cycles.

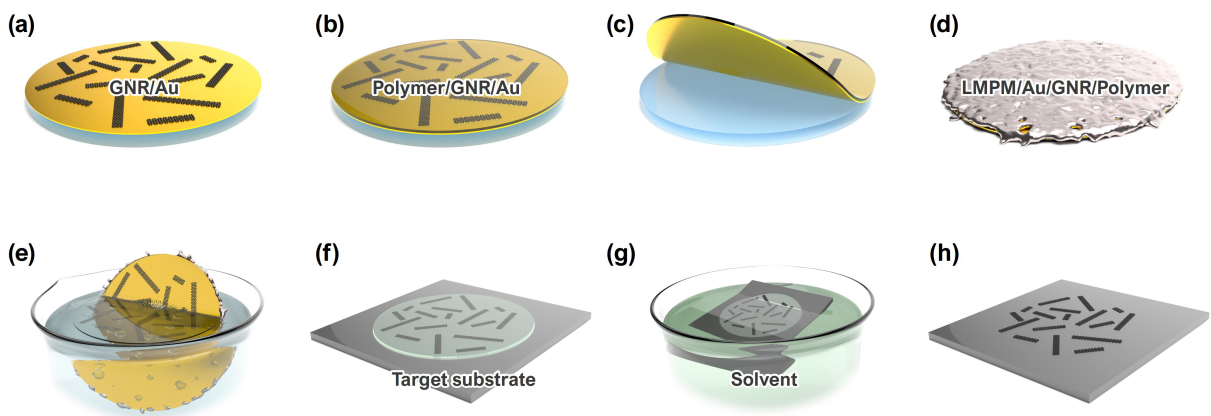

Figure 1: Schematic illustration of the LMPM-assisted transfer procedure. (a) Deposition of a metal thin film onto the initial substrate, followed by the growth of low-dimensional materials on top of the metal film. (b) Spin-coating of a polymer film onto the low-dimensional materials and metal thin film. (c) Detachment of the polymer/metal film stack from the substrate. (d) Application of liquid Field's metal (LMPM) to the metal film side, which is then solidified. (e) Hydrogen bubble delamination at the polymer/metal interface, enabling clean separation without damaging the materials. (f) Transfer of low-dimensional materials along with the polymer layer onto a target substrate. (g) Removal of the polymer layer using a solvent. (h) Low-dimensional materials transfer completed.

To demonstrate this process in practice, we first performed LMPM-assisted transfer of 7-armchair graphene nanoribbons (7-AGNRs). The GNRs were grown on an atomically smooth Au thin film deposited on mica substrate using the well-established on-surface synthesis method employing 10,10'-dibromo-9,9'-bianthracene (DBBA) [7], as shown in Figure 2a. The Au(111) surface was confirmed by X-ray diffraction (XRD) (Figure S1). The growth details are provided in the Experimental Methods section.

In addition to Raman spectroscopy confirmation discussed later, *ex situ* scanning tunneling microscopy (STM) was conducted to verify the successful growth of GNRs. Figures 2b and c show STM topographic images of the 7-AGNRs at low and high magnification, respectively. The images reveal distinct ribbon-like features with smooth edges, with an average length of ~15 nm and width of ~0.9 nm. The overlaid chemical structure of 7-AGNR matches the observed features well. Both the measured dimensions and the overlays confirm the expected geometry and identity of 7-AGNRs. We also observe the presence of possible double-layer stacking in Figure 2c, which has been previously reported in the literature [47–49]. The sample handling after growth and the details of the STM imaging are provided in the Experimental Methods section. STM images of several other regions with and without GNRs are included in the Supporting Information (Figure S2).

Figure 2d-i shows photographs of the transfer process, including liquid LMPM application, hydrogen bubble delamination, and final placement of GNRs onto a target substrate. To evaluate the structural integrity and uniformity of the ribbons before and after transfer, Raman spectra were collected using a 532 nm laser. This wavelength was selected to match the $E_{22}$ optical transition energy of 7-AGNRs, enhancing Raman intensity [46]. Figure 2j shows the Raman spectra of 7-

AGNRs pre-transfer and post-transfer. Pre-transfer, Raman spectra exhibited characteristic peaks, including the G peak (~1599 cm$^{-1}$), D peak (~1342 cm$^{-1}$), C-H bending modes (~1222–1262 cm$^{-1}$), and radial-breathing-like mode (RBLM, ~399 cm$^{-1}$), which are consistent with values reported in the literature [7,50–52]. The sharpness of all peaks reflects the structural uniformity of the synthesized ribbons.

The single intense RBLM peak serves as a fingerprint for GNRs [53], indicating their structural homogeneity. The full-range Raman spectra covering the entire Raman shift range shown in Figure 2j are provided in the Supporting Information for both pre- and post-transfer (Figure S3).

Post-transfer Raman analysis confirmed that all characteristic peaks were retained without significant change, demonstrating that the LMPM-assisted method maintains GNR structural integrity (Figure 2j). No significant changes were observed in the width or position of low-frequency modes such as RBLM after transfer, while a slight upshift in the G peak was detected. This shift is likely due to interactions with the SiO$_2$/Si substrate rather than intrinsic defects in the ribbons, consistent with previous studies [46,47]. To further confirm these observations, peak fitting analysis was performed on more than 100 Raman spectra, and statistical evaluations of peak positions and widths showed consistent results. Detailed results are presented in Figures S3-4 and Table S1. We also applied this transfer method to 9-armchair graphene nanoribbons (9-AGNRs) synthesized under similar conditions and confirmed that the structural integrity was maintained during transfer. The Raman spectra of 9-AGNRs before and after transfer, showing their characteristic peaks, are presented in Figure S5.

To assess the cleanliness of the transferred GNRs, X-ray photoelectron spectroscopy (XPS) was performed to detect possible metal residues on the surface, and the results were compared with those obtained using a chemical approach based on Au etching. Typically, the detachment of the GNR/Au film from the mica substrate in conventional Au etching transfer methods relies on HCl treatment [44–46], which requires prolonged processing times. To address this issue, we developed an HCl-free procedure that enables quick transfer. This HCl-free procedure is implemented through a modified Au etching method, illustrated in Figure S6 and described in the Experimental Methods section.

No detectable peaks corresponding to LMPM elements were observed in the XPS spectra of samples transferred by the LMPM-assisted method, indicating that any residues of these elements are below the detection limit of XPS. Similarly, no detectable peaks for LMPM elements were observed in samples transferred using the modified Au etching procedure. Regarding Au residues, the Au etching method showed no detectable Au peaks, while the LMPM-assisted method exhibited only extremely weak Au peaks, barely above the detection limit. These negligible Au signals confirm that both methods achieve comparable cleanliness in terms of metal contamination. The corresponding XPS spectra are shown in Figures S7-11. These results demonstrate that the LMPM-assisted transfer achieves a level of cleanliness comparable to the conventional chemical transfer approach using Au etching, but without requiring metal etchants.

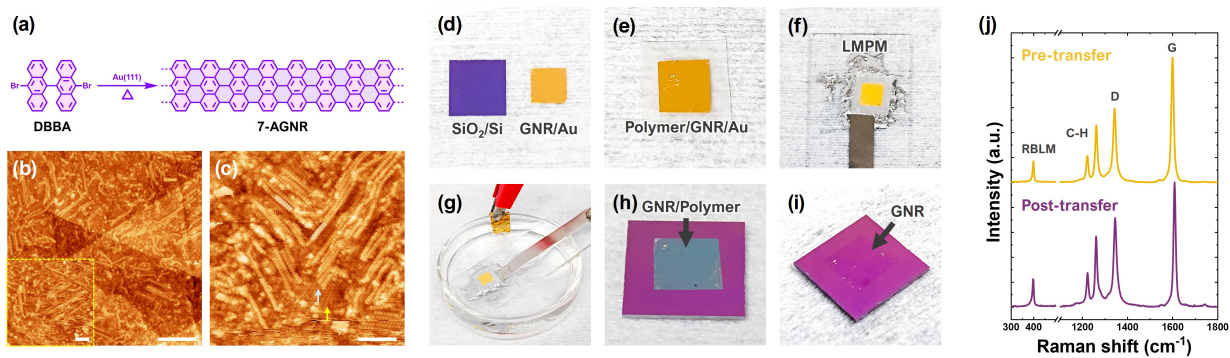

Figure 2: (a) On-surface synthesis scheme for 7-AGNR. (b-c) STM topographic images of 7-AGNRs on Au(111) on mica at 77 K, (Vs = 1.5 V, I = 100 pA); scale bar: b = 20.0 nm, inset b, c = 10.0 nm. Yellow arrow indicates monolayer and blue arrow indicates double layer. (d-i) Photographs and Raman analysis of the LMPM-assisted transfer process for GNRs: (d) Target substrate ($SiO_2$/Si) and 7-AGNRs grown on Au/mica. (e) 7-AGNRs/Au with spin-coated polymer film peeled off from the mica substrate. (f) Application of liquid LMPM to the Au film side. (g) Hydrogen bubble delamination at the polymer/Au interface. (h) Placement of GNRs onto the target substrate. (i) GNRs left on the target substrate after polymer removal by solvent. (j) Raman spectra of 7-AGNRs pre- (yellow) and post-transfer (purple), confirming structural integrity and minimal defect density.

While GNR transfer onto $SiO_2$/Si substrate was demonstrated, we further explored the versatility of our LMPM-assisted method by transferring 7-AGNRs onto two-dimensional (2D) van der Waals dielectric materials. Unlike conventional dielectrics fabricated by commonly used atomic layer deposition (ALD) techniques, which inevitably introduce surface roughness, 2D dielectrics maintain their pristine flatness, a critical advantage for next-generation nanoelectronic devices that require atomically precise interfaces. For this demonstration, we selected LaOCl as an exemplary 2D dielectric, a promising material with high dielectric constant and high breakdown voltage, as its atomically flat surfaces provide an ideal platform for evaluating the quality and integrity of transferred materials [54–56].

Using the same LMPM-assisted method, we transferred the 7-AGNRs from Au film to LaOCl. LaOCl crystals were synthesized on a mica substrate using a molten salt method [55]. The details of the synthesis are described in the Experimental Methods section. Stoichiometry and crystal structure of synthesized crystals were confirmed by energy dispersive spectroscopy (EDS), X-ray diffraction (XRD) and Raman spectroscopy (Figure S12).

High-Angle Annular Dark Field Scanning Transmission Electron Microscopy (HAADF-STEM) was used to demonstrate the 2D structure of LaOCl flakes as expected from the synthesis method, shown in Figure 3a-b. Figure 3c further shows the atomic resolution HAADF-STEM image of the LaOCl oriented along the low index (001) zone axis corroborating the crystallinity of the material. Additionally, the surface morphology was measured by atomic force microscopy (AFM) of LaOCl crystals before and after 7-aGNR transfer to evaluate the physical uniformity and topographical continuity of the transferred ribbons (Figure 3d-e). Pristine LaOCl crystals have atomically smooth surfaces as shown in Figure 3d. While the surface after transfer has some polymer residues, the clean area shows a small roughness of ~0.3 nm (Figure 3e). Polymer residue is a common concern in polymer-assisted transfer methods. However, recent literature has reported significant progress in reducing polymer residues through careful selection of polymer types and optimization of processing conditions, suggesting that further improvements to minimize residues are possible [6,57,58]. This suggests that the LMPM-assisted transfer method preserves both the intrinsic structures as confirmed by Raman and the microscopic physical integrity of the materials during transfer.

The transfer and molecular structural preservation of 7-AGNRs before and after transfer were confirmed by Raman spectroscopy (Figure 3f). We observed the characteristic Raman peaks of 7-AGNRs after transfer: the G peak (~1609 cm$^{-1}$), D peak (~1344 cm$^{-1}$), C-H bending modes (~1224–1269 cm$^{-1}$). Additionally, Raman modes corresponding to LaOCl were detected at ~333 cm$^{-1}$ ($B_{1g}$) and ~437 cm$^{-1}$ ($E_g$), which are consistent with previously reported values in the literature [55,56,59–61]. Several Raman peaks overlapped with those of mica or 7-AGNRs. These results confirm the successful transfer of 7-AGNRs onto LaOCl without degradation of their intrinsic structures.

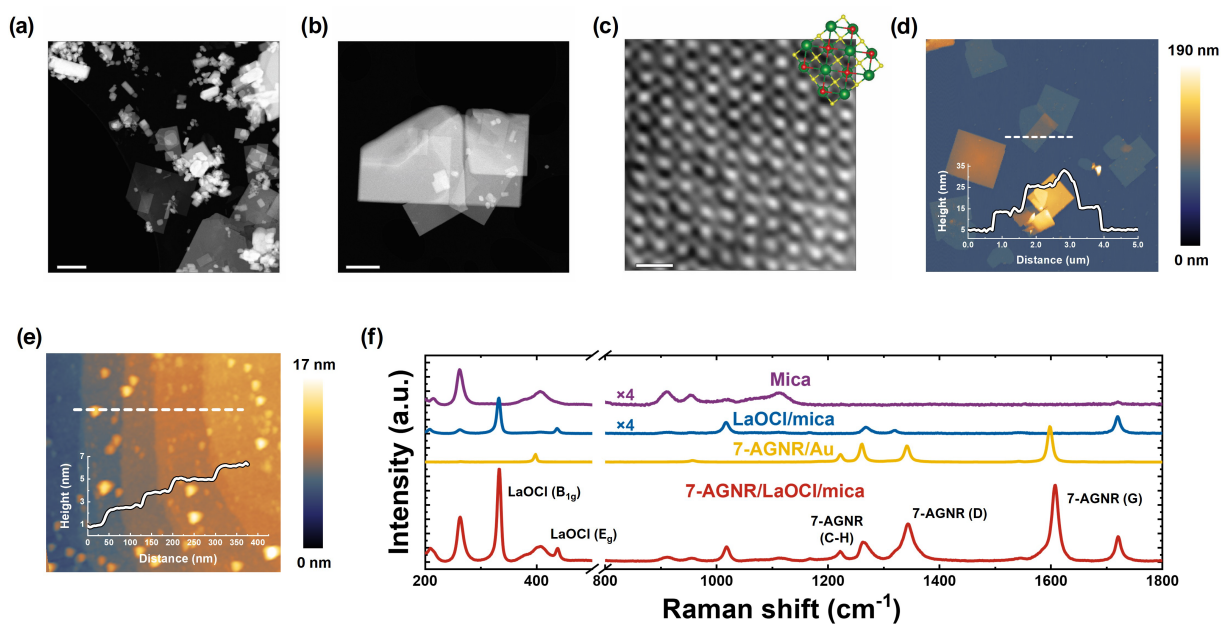

Figure 3: (a) Low magnification HAADF-STEM image of LaOCl showing several flakes with varying 2D dimensions. Scale bar is 1 μm. (b) High magnification HAADF-STEM image of a few stacked 2D LaOCl flakes. Scale bar is 1 μm. (c) Atomic resolution HAADF-STEM image of LaOCl shows the (001) zone axis orientation. The image is overlaid with the structural model of LaOCl showing La (green spheres), O (yellow spheres), and Cl (red spheres). Scale bar is 0.5 nm. (d) AFM image of LaOCl crystals before 7-AGNR transfer. (e) AFM image of LaOCl crystals after 7-AGNR transfer. (f) Raman spectra of substrate (mica), LaOCl, and 7-AGNR before and after transfer onto LaOCl.

To demonstrate the scalability of the LMPM-assisted transfer method, we performed a wafer-scale transfer using an Au film sputter-deposited onto a 100 mm sapphire wafer. Due to the size limitations of our synthesis system, GNRs were not grown on the Au film, and the transfer process was demonstrated without synthesis. However, since the transfer mechanism remains identical to that described in the previous sections, this wafer-scale procedure is expected to be applicable for GNR transfer as well.

The wafer-scale transfer process, depicted in Figure 4, follows the same steps as those used for GNR transfer described in the Experimental Methods section, except that no GNRs were grown on the Au film. The process includes sputtering a 600 nm Au film onto a sapphire substrate, applying liquid LMPM, and performing hydrogen bubble delamination to remove the polymer layer. For detailed procedures, refer to Figure 4 and the Experimental Methods section. It is difficult to spread LMPM over a large area on certain surfaces like glass due to its low wettability, as can be seen in Figure 2f. In this demonstration, we applied LMPM directly onto the Au film surface to address this issue. This limitation can be overcome by using suitable substrates with better wettability for LMPM, such as PDMS, as demonstrated in Figure S13.

As shown in Figure 4e and f, the delaminated polymer film retained its wafer-scale dimensions without any visible damage, demonstrating the robustness and reliability of this method for large-area transfers. This result highlights the potential of LMPM-assisted transfer for scaling up low-dimensional materials integration.

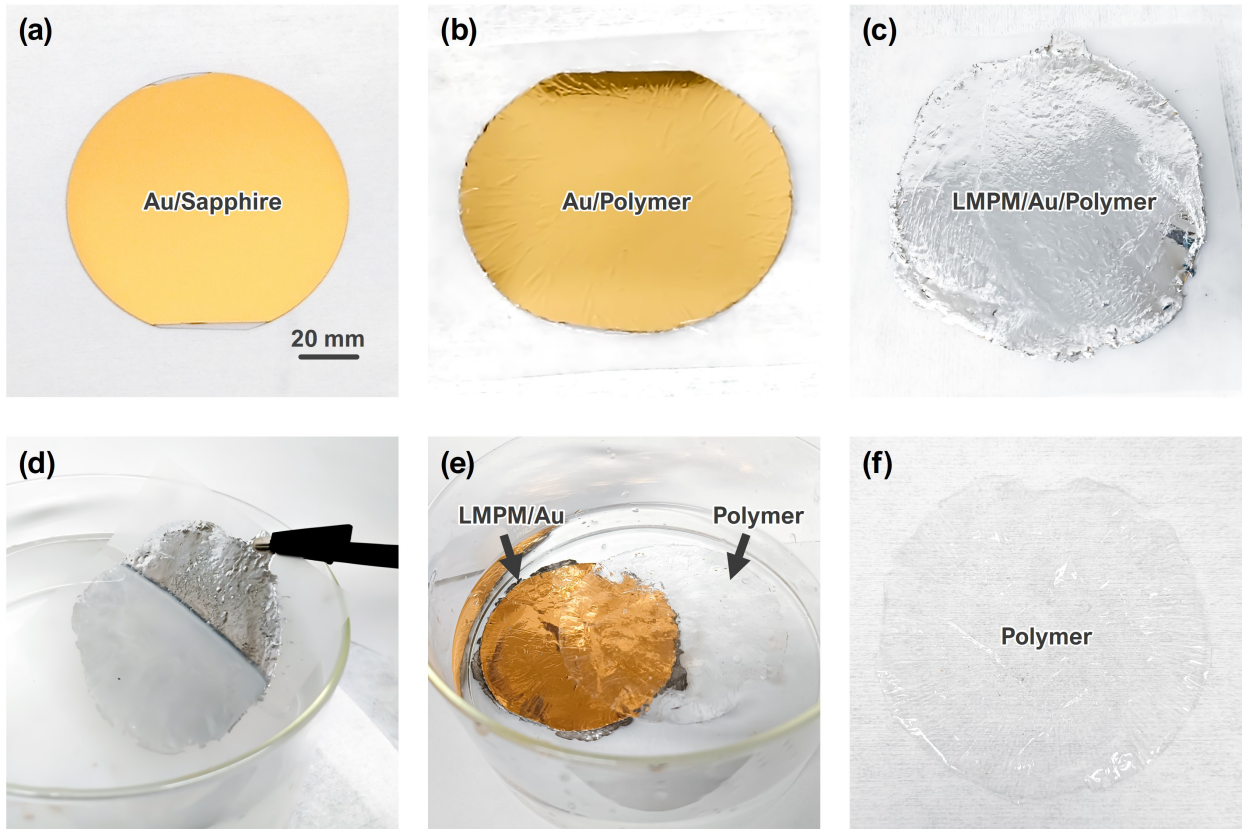

Figure 4: Wafer-scale transfer process using LMPM-assisted method. (a) Au film sputter-deposited on a 100 mm sapphire wafer. (b) Polymer-coated Au film peeled off from the sapphire substrate. (c) Application of melted LMPM to the Au side. (d) Hydrogen bubble delamination in NaOH solution to separate the polymer layer. (e) Polymer layer successfully separated from Au and LMPM layers. (f) Photograph of the delaminated polymer film, demonstrating wafer-scale transfer without damage.

Finally, we fabricated field-effect transistor (FET) devices using 9-AGNRs transferred via our etch-free method. The illustration of the device structure and the scanning electron microscopy (SEM) image of the fabricated device are shown in Figure 5a-b. First, 9-AGNRs were transferred onto pre-patterned, locally backgated device substrates, then source and drain electrodes were fabricated on the transferred 9-AGNRs. The Raman spectrum of the ribbons after transfer onto the device maintains the sharp characteristic peaks of 9-AGNRs as shown in Figure S14, confirming the structural integrity of the transferred GNRs. The details of the fabrication of the pre-patterned device structure and the source and drain electrodes are described in the Experimental Methods section.

A representative $I_D$-$V_{GS}$ characteristic of the fabricated 9-AGNR FETs is shown in Figure 5c. The fabricated devices exhibit p-type switching behavior, the same as previously reported FETs with 9-AGNRs transferred by conventional methods[62–64], with on/off current ratios reaching ~1000. This demonstrates that our approach can be an effective alternative for device integration. Figure 5d shows nonlinear $I_D$-$V_{DS}$ characteristics indicative of Schottky barrier-limited transport [62,65,66]. This limitation may be attributed to Ti/Au's low work function, which can lead to higher Schottky barriers at the metal–GNR interface. Similar barriers were also observed in devices fabricated via conventional methods and can be improved by optimizing metal selection and refining fabrication processes to reduce defects [62]. Overall, the demonstrated transistor functionality validates the feasibility of our LMPM-assisted transfer method for integrating low-dimensional materials in practical electronic applications.

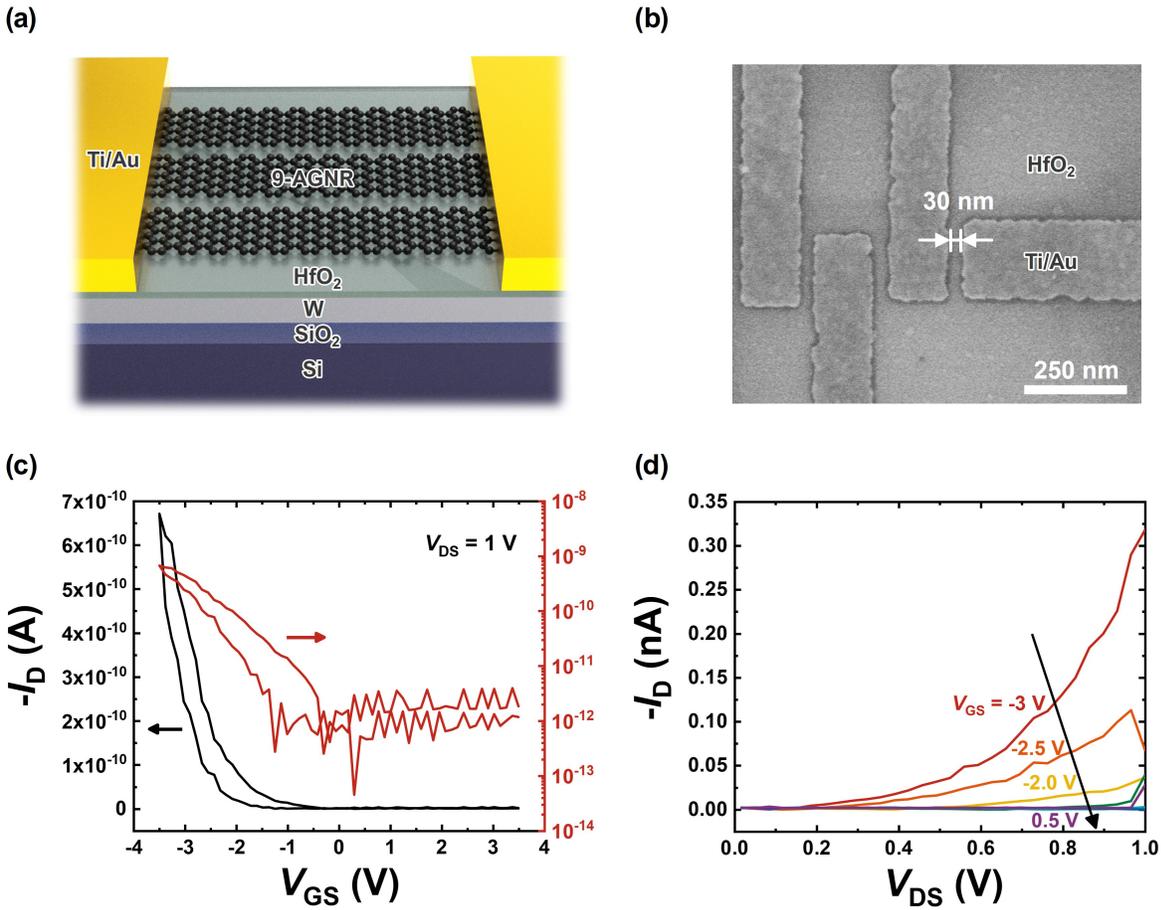

Figure 5: Fabrication and electrical transport characterization of 9-AGNR FETs. (a) Schematic of 9-AGNR FET device structure. (b) SEM image of the fabricated FET device electrodes. (c) $I_D$-$V_{GS}$ characteristics of 9-AGNR FET in linear scale (black line) on the left axis and log scale (red line) on the right axis. Data taken with $V_{DS}$ = 1 V. (d) $I_D$-$V_{DS}$ characteristics of 9-AGNR FET. Measurements were performed with $V_{GS}$ ranging from -3 V to 0.5 V with 0.5 V increments.

## Conclusions

In this work, we have developed a scalable, etch-free transfer method for low-dimensional materials using low-melting-point metals (LMPMs) as mechanical support. This approach addresses limitations of conventional transfer techniques while maintaining material integrity and quality. Using Field's metal (51% In, 32.5% Bi, 16.5% Sn by weight) with a low melting point of 62°C, we transferred 7-armchair graphene nanoribbons (7-AGNRs) from Au thin films to various substrates including $SiO_2$/Si and 2D van der Waals dielectric materials (LaOCl) without chemical etching. Raman spectroscopy confirmed structural uniformity and minimal defect density after transfer, with all characteristic peaks retained. Furthermore, we demonstrated the scalability of our method using 100-mm diameter Au thin films on sapphire substrates. The LMPM-assisted transfer method provides significant improvements including metal film reusability, reduced chemical waste, low-temperature processing compatibility, and recyclability of support materials. We also fabricated FET devices using transferred GNRs, which exhibited room-temperature switching behavior with on/off current ratios reaching ~1000. These device results validate the compatibility of our transfer method with functional device integration. Altogether, this LMPM-assisted method represents a versatile, reusable, and environmentally conscious platform for transferring low-dimensional materials. It could motivate further investigation into integration strategies for functional nanomaterials and facilitate the incorporation of these materials into large-scale, practical device applications across diverse technological domains.

## Experimental Methods

**Synthesis of Graphene Nanoribbons**

Precursor monomers for 7-AGNR and 9-AGNR were synthesized using previously reported procedures [7,67]. The growth was performed using a fully automated in-house ultra-high vacuum (UHV) system dedicated to GNR synthesis (Createc MiniMBE System Type RS2-M-4-FS). Au/mica substrates (Phasis, Switzerland) were cleaned in an UHV chamber by sputtering with 1 kV $Ar^+$ ions for 1 hour, followed by annealing at 470 °C for 1 hour. For the growth of 9-AGNRs, the precursor monomer 10,10'-dibromo-9,9'-bianthryl (DBTP) was sublimated onto the clean Au surface from a quartz crucible heated to 105 °C, while the substrate temperature was maintained at 200 °C. Molecule deposition was carried out for 10 minutes. After deposition, the substrate was heated to 400 °C at a rate of 0.5 K/s and held for 10 minutes to induce cyclodehydrogenation, forming the GNRs.

For the growth of 7-AGNRs, the precursor monomer 10,10'-dibromo-9,9'-bianthracene (DBBA) was sublimated onto the clean Au surface from a quartz crucible heated to 230 °C, with the substrate temperature similarly maintained at 200 °C. Molecule deposition was performed for 30 minutes. Following deposition, the substrate was heated to 400 °C at a rate of 0.5 K/s and held for 10 minutes to complete cyclodehydrogenation and form the GNRs.

**Scanning Tunneling Microscopy (STM) of 7-AGNR**

The experiment was performed using a low-temperature scanning tunneling microscope (Createc). Chemically etched tungsten tips were used for imaging the sample. The *ex situ* 7-AGNR films on Au(111) on mica were first degassed under UHV at ~ 5 ×$10^{-8}$ mbar pressure in the load lock chamber for 24 hours before transferring to the STM chamber with a base pressure of ~ $4 \times 10^{-10}$ mbar. Then the samples were annealed at 300 °C for 3 hours to reduce ambient contamination before imaging.

**Low-Melting-Point Metal-Assisted Transfer of Graphene Nanoribbons**

7-AGNRs were synthesized on Au/mica substrates. A polycarbonate (PC) film was spin-coated onto the 7-AGNR/Au/mica surface and cured at 80 °C for 10 minutes. The assembly (PC layer, GNRs, and Au film) was detached from the mica substrate by immersing the sample in water.

Field's metal (51% In, 32.5% Bi, 16.5% Sn by weight) was melted at 70 °C and applied to the Au side of the PC/GNR/Au assembly. After cooling, the assembly underwent electrochemical delamination in a 1 M NaOH aqueous solution. The Field's metal-covered Au film served as the cathode and an Au foil as the anode. A DC voltage of 5 V was applied, generating hydrogen bubbles between the Au film and PC/GNRs via water reduction ($2H_2O + 2e^- \rightarrow H_2 + 2OH^-$). These bubbles separated the PC/GNR layer from the Au film.

The delaminated PC/GNR layer was rinsed with water and transferred onto a target substrate. The sample was annealed at 160 °C for 10 minutes to improve adhesion. The PC support layer was then removed by sequential rinsing with chloroform, acetone, isopropyl alcohol, and water.

**Transfer of Graphene Nanoribbons Using Au Etchant**

In conventional transfer methods, HCl is used to detach the GNR/Au film from the mica substrate [44–46]. However, HCl and other acids can potentially alter the properties of carbon materials and require lengthy processing times [68]. To address these issues, we developed an HCl-free procedure that enables quick transfer. A schematic illustration of the transfer process is provided in Figure S6 of the Supporting Information. 7-AGNRs were synthesized on Au/mica substrates. A hole corresponding to the desired transfer shape was cut into Scotch tape. The GNR/Au film was picked up from the mica using the Scotch tape with the hole. The Scotch tape/GNR/Au stack was flipped upside down and placed onto the target substrate. The edges of the suspended Au film were trimmed to separate the Au film from the Scotch tape. The Scotch tape was removed, leaving the detached GNR/Au film on the target substrate. Finally, the Au layer was etched away using a potassium iodide (KI) solution, leaving only the GNRs on the substrate.

**Synthesis of LaOCl Crystals**

LaOCl nanoflakes were synthesized using a molten salt method, as previously reported [55]. $LaCl_3 \cdot xH_2O$ was ground with a KCl-LiCl eutectic mixture (41.8 mol% KCl) in a mortar. The mixture was dispersed onto freshly cleaved mica and heated to 500°C in a furnace under air atmosphere. After cooling, the LaOCl nanoflakes were obtained by washing with deionized water to remove residual salts.

## Raman Spectroscopy

Raman measurements were performed using a Renishaw Raman microscope with a 785 nm laser for 9-AGNRs and a Horiba Raman microscope with a 532 nm laser for 7-AGNRs. The laser power was kept below 10 mW, and a 50× objective lens was used in both cases. For a statistical analysis study, over 100 spectra were collected from different points on each sample of 7-AGNRs.

## Scanning Electron Microscopy Characterization

SEM imaging was carried out on a Hitachi S-4800 system.

## Preparation of Pre-Patterned Local Bottom Gate Chips

The pre-patterned devices were fabricated on substrates consisting of 100 nm thermally grown $SiO_2$ on heavily doped Si wafers. A local back-gate electrode was first defined by sputtering an ~8 nm tungsten (W) layer, which was patterned using photolithography and selectively etched with hydrogen peroxide ($H_2O_2$). Subsequently, a ~5.5 nm hafnium oxide ($HfO_x$) gate dielectric was deposited by atomic layer deposition (ALD) at 135 °C using tetrakis(dimethylamino)hafnium ($C_8H_{24}HfN_4$) as a precursor. Alignment marks and contact pads were fabricated via photolithography and lift-off process, using a ~3 nm chromium (Cr) adhesion layer and ~25 nm platinum (Pt) layer. The processed wafers were then diced into individual chips, each measuring 1 cm × 1 cm and containing over 250 local back-gated devices, with each device region housing six individual transistors

## Contact Metal Fabrication

Source and drain contact electrodes were fabricated using an Elionix ELS-7000 electron beam lithography system followed by metal deposition with a Temescal FC-2500 e-beam evaporator. Following the transfer of 9-AGNRs, the device substrates were spin-coated with a bilayer e-beam resist system. The first layer consisted of MMA EL6, spin-coated at 4000 rpm for 1 minute and baked at 150 °C for 5 minutes. The second layer of PMMA 950K A2 was spin-coated at 4500 rpm for 1 minute and baked at 180 °C for 10 minutes. Electron beam patterning was performed using a 50 pA beam current. Following development, Ti/Au (3/12 nm) was deposited to form the source and drain electrodes. The channel lengths of the resulting devices were between 30-35 nm, with electrode widths and lengths ranging from 150-200 nm and 150-500 nm, respectively.

## Electrical Transport Measurements

The electrical transport measurements were performed using a Lakeshore TTPX cryogenic probe station, in conjunction with M81-SSM synchronous source measure system and MeasureLINK software. All measurements were conducted under ambient conditions unless specified.

## Scanning Transmission Electron Microscopy

The specimens were investigated with an FEI Titan G2 60-300 (S)TEM microscope equipped with a CEOS DCOR probe corrector, a monochromator, and a super-X EDX spectrometer. The microscope was operated at 200 kV with a STEM incident probe convergence angle of 25.5 mrad and a probe current of 25 pA. HAADF-STEM images were acquired at a detector inner and outer collection angle of 55 and 200 mrad respectively.

## Acknowledgments

K.Y., M.Y., and M.S. contributed equally to this work. This research was primarily supported by grants from Semiconductor Research Corporation (SRC) (#3144.001) for device fabrication and characterization, and National Science Foundation (NSF) (#2235143) for GNR synthesis, characterization and processing. Partial support was also provided by Arizona Technology and Research Initiative Fund (TRIF) for major equipment acquisition. F.D. and S.J. were supported through fellowships from SRC Undergraduate Research Program (URP). R.H. acknowledges support from the University of Arizona (UA) Vertically Integrated Projects (VIP) program. H.Y. and T.S. acknowledges partial support from NSF STC (#2242925). ALD was performed at Nanofabrication Facility within Molecular Foundry Cleanroom at Lawrence Berkeley National Laboratory (LBNL), supported by the U.S. Department of Energy, Office of Science, and Office of Basic Energy Sciences (# DE-AC02-05CH11231). We acknowledge the funding for instrumentation in the Kuiper-Arizona Laboratory for Astromaterials Analysis at the UA (Raman spectroscopy and SEM characterization) from NASA grants #80NSSC23K0327, #NNX12AL47G, #NNX15AJ22G, and #NNX07AI520, and NSF grants #1531243 and #EAR-0841669. The electron microscopy portion of this work by R.R and K.A.M. was supported by NSF through award No. DMR-2309431, carried out in the Characterization Facility of the University of Minnesota, which receives partial support from the NSF through the MRSEC program under award DMR-2011401. All X-ray and Ultraviolet Photoelectron Spectra were collected at the Laboratory for Electron Spectroscopy and Surface Analysis (LESSA) in the Department of Chemistry and Biochemistry at the University of Arizona using a Kratos Axis 165 Ultra DLD Hybrid Ultrahigh Vacuum Photoelectron Spectrometer. The instrument was purchased with funding from the National Science Foundation and supported by the Center for Interface Science: Solar-Electric Materials (CIS:SEM), an Energy Frontier Research Center funded by the U.S. Department of Energy and Arizona Technology and Research Initiative Fund (A.R.S.§15-1648). All Park Systems NX20 AFM images and data were collected in the W.M. Keck Center for Nano-Scale Imaging, RRID:SCR_022884, in the Department of Chemistry and Biochemistry at the University of Arizona. This instrument purchase was partially supported by Arizona Technology and Research Initiative Fund (A.R.S.§15-1648). The work on GNR precursor synthesis and STM characterization of ribbons was supported by the Office of Naval Research, award N00014-19-1-2596 (M.S. and A.S.).

## Author Contributions

K.Y. and Z.M. conceptualized the work. Z.M. supervised the project. M.S., under the supervision of A.S., synthesized the precursor molecules of GNRs. K.Y. and M.Y. synthesized and characterized the GNRs with support from A.C.W.. K.Y. synthesized and characterized the LaOCl crystals with support from P.D.. F.D. and R.H. prepared the Au thin films for wafer-scale transfer. J.B. characterized the Au films used for GNR synthesis. M.S., under the supervision of A.S., performed the scanning tunneling microscopy imaging of the synthesized GNRs. R.R., under the supervision of K.A.M., conducted the scanning transmission electron microscopy imaging of LaOCl. M.Y. fabricated the devices with support from H.Y. and T.S., and conducted the electrical measurements with support from S.J.. K.Y. led the writing of the manuscript and the preparation of figures, with contributions from M.Y., M.S., and R.R. All authors provided feedback on the manuscript. Z.M., A.S., and K.A.M. were responsible for funding acquisition. Z.M. finalized the manuscript.

# Scalable Etch-Free Transfer of Low-Dimensional Materials from Metal Films to Diverse Substrates

*Kentaro Yumigeta [a†], Muhammed Yusufoglu [a†], Mamun Sarker [b†], Rishi Raj [c], Franco Daluisio [d], Richard Holloway [a], Howard Yawit [a], Thomas Sweepe [a], Julian Battaglia [a], Shelby Janssen [a], Alex C. Welch [d], Paul DiPasquale [a], K. Andre Mkhoyan [c], Alexander Sinitskii [b], Zafer Mutlu [a,e,f,]\**

[a] Department of Materials Science & Engineering, University of Arizona, Tucson, Arizona 85721, USA

[b] Department of Chemistry, University of Nebraska-Lincoln, Lincoln, Nebraska 68588, USA

[c] Department of Chemical Engineering and Materials Science, University of Minnesota, Minneapolis, MN 55455, USA

[d] Department of Chemical and Environmental Engineering, University of Arizona, Tucson, Arizona 85721, USA

[e] Department of Electrical and Computer Engineering, University of Arizona, Tucson, Arizona 85721, USA

[f] Department of Physics, University of Arizona, Tucson, Arizona 85721, USA

† These authors contributed equally to this work.

\* Corresponding Author: zmutlu@arizona.edu

**XRD Analysis of Crystallographic Orientation in Au Films on Mica Substrates**

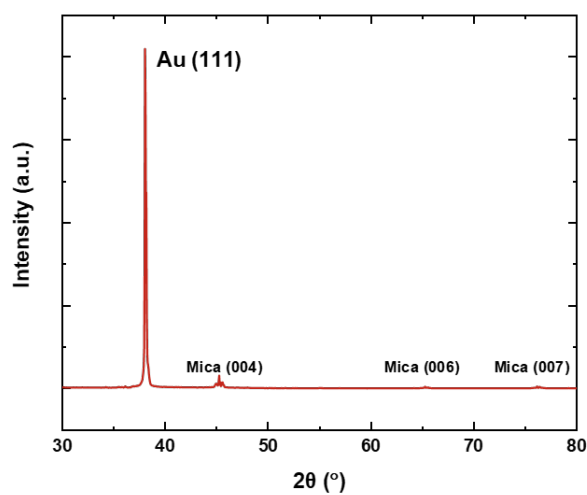

Figure S1: XRD spectrum of Au/mica substrates.

**Scanning Tunneling Microscopy (STM) of the Areas with and without 7-AGNRs**

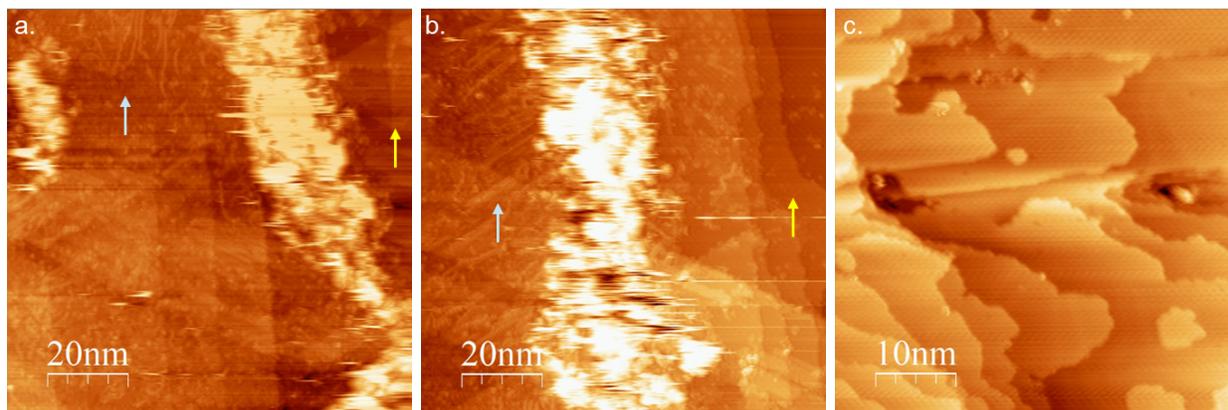

Figure S2: Additional STM topographic images of 7-AGNRs on Au(111) on mica at 77 K, (Vs = 1.5 V, I = 100 pA). Blue arrows indicate areas with GNRs and yellow arrows indicate ribbon-free areas.

The images reveal surface contamination, which appears to preferentially accumulate in regions where GNRs are present. Areas without GNRs appear cleaner, suggesting that contamination, likely introduced during air exposure prior to *ex situ* STM imaging, tends to localize around the ribbons.

**Raman Spectroscopy Analysis of 7-AGNRs Pre- and Post-Transfer**

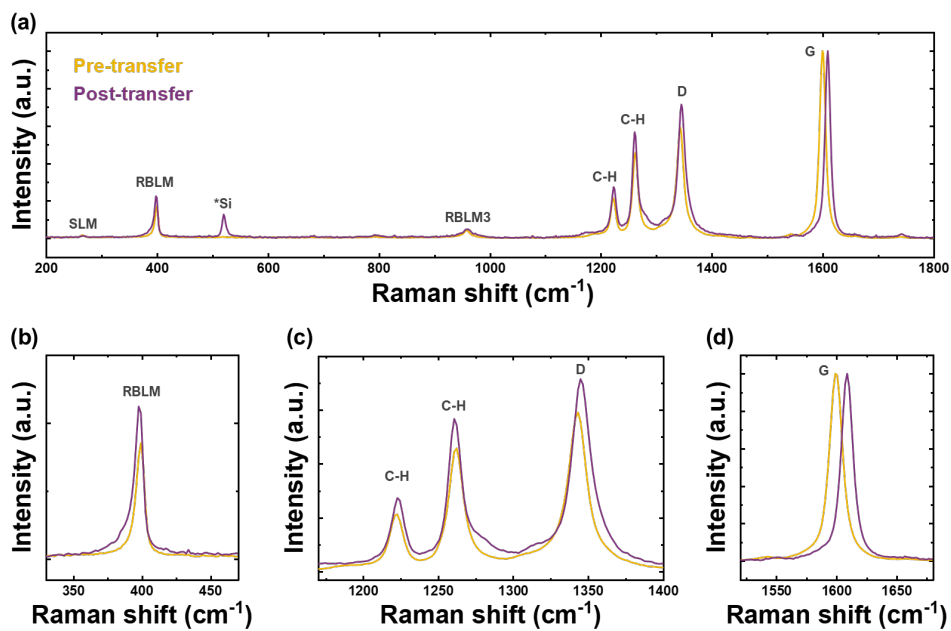

Figure S3: Raman spectra of graphene nanoribbons (7-AGNRs) before and after transfer using the low-melting-point metal (LMPM)-assisted method. (a) Full-range Raman spectra of GNRs before (yellow) and after (purple) transfer. The spectra are normalized to the intensity of the G peak. Key characteristic peaks, including the radial breathing-like mode (RBLM), D peak, C-H bending modes, and G peak, are preserved after transfer, indicating minimal structural or chemical modification. (b-d) Magnified views of specific Raman regions: (b) RBLM, (c) C-H bending modes and D peak, and (d) G peak.

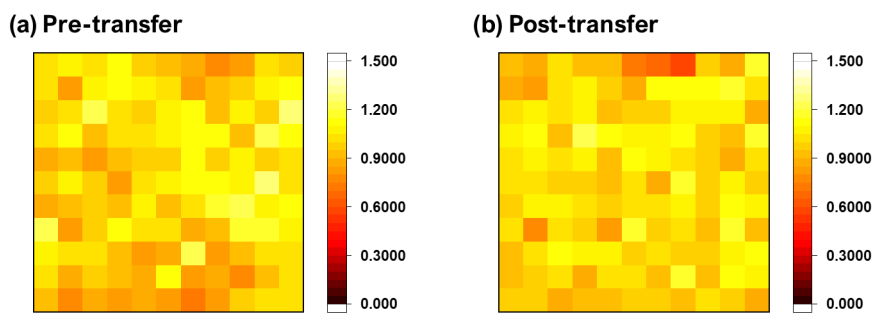

Figure S4: Raman mapping of the G peak intensity for 7-AGNRs before and after LMPM-assisted transfer. Raman mapping was performed over a 100 × 100 μm area to evaluate the uniformity of the G peak intensity. (a) Mappings pre-transfer and (b) mapping post-transfer demonstrate consistent intensity distribution, indicating that the LMPM-assisted transfer process preserves the spatial uniformity of the GNRs. No significant changes in intensity variation or dispersion were observed, confirming the robustness of the transfer method.

Table S1: Summary of Raman peak fitting results for 7-AGNRs pre- and post-transfer. The table compares the normalized intensity (calculated relative to the mean peak intensity), center position, and full width at half maximum (FWHM) of key Raman peaks (G, C-H, D, and RBLM) for 7-AGNRs. All values were calculated based on fitting results from over 100 Raman spectra collected pre- and post-transfer.

|  |  | Before transfer | After transfer |
|---|---|---|---|
| G | Normalized Intensity (a.u.) | 1.00 ± 0.12 | 1.00 ± 0.10 |
|   | Center (cm$^{-1}$) | 1599.48 ± 0.08 | 1608.68 ± 0.11 |
|   | FWHM (cm$^{-1}$) | 12.44 ± 0.21 | 11.02 ± 0.34 |
| C-H | Normalized Intensity (a.u.) | 1.00 ± 0.14 | 1.00 ± 0.16 |
|   | Center (cm$^{-1}$) | 1221.96 ± 0.13 | 1222.95 ± 0.26 |
|   | FWHM (cm$^{-1}$) | 9.16 ± 0.29 | 8.69 ± 0.67 |
| C-H | Normalized Intensity (a.u.) | 1.00 ± 0.13 | 1.00 ± 0.11 |
|   | Center (cm$^{-1}$) | 1261.93 ± 0.10 | 1261.33 ± 0.16 |
|   | FWHM (cm$^{-1}$) | 11.23 ± 0.26 | 10.62 ± 0.57 |
| D | Normalized Intensity (a.u.) | 1.00 ± 0.11 | 1.00 ± 0.09 |
|   | Center (cm$^{-1}$) | 1342.70 ± 0.08 | 1345.30 ± 0.16 |
|   | FWHM (cm$^{-1}$) | 15.26 ± 0.39 | 15.52 ± 0.51 |
| RBLM | Normalized Intensity (a.u.) | 1 ± 0.14 | 1 ± 0.13 |
|   | Center (cm$^{-1}$) | 398.79 ± 0.19 | 397.71 ± 0.30 |
|   | FWHM (cm$^{-1}$) | 6.41 ± 0.46 | 6.84 ± 0.86 |
| $I_D/I_G$ |  | 0.6634±0.1051 | 0.9319±0.1092 |

**Raman Spectroscopy Analysis of 9-AGNRs Pre- and Post-Transfer**

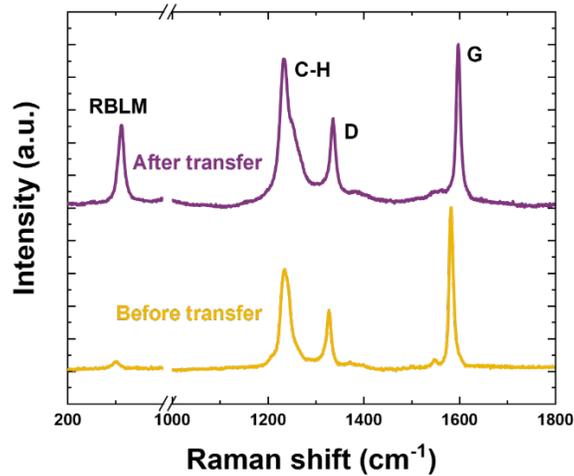

Figure S5: Raman spectra of graphene nanoribbons (9-AGNRs) before and after transfer from Au/mica onto SiO$_2$/Si substrates using the low-melting-point metal (LMPM)-assisted method.

**Au Etchant-Based HCl-Free Transfer Process of Graphene Nanoribbons**

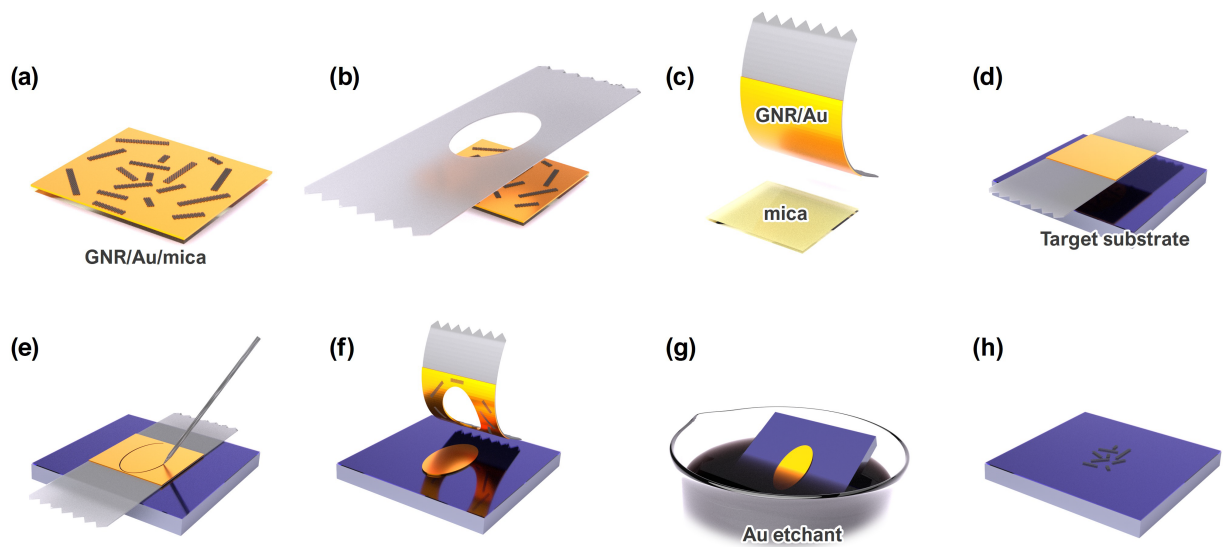

Figure S6: Step-by-step schematic illustration of the transfer process for 7-AGNRs using Au etching: (a) 7-AGNRs are synthesized on Au/mica substrates. (b) A hole corresponding to the desired transfer shape is cut into Scotch tape. (c) The GNR/Au film is picked up from the mica using Scotch tape with the hole. (d) The Scotch tape/GNR/Au stack is flipped upside down and placed onto the target substrate. (e) The edges of the suspended Au film are trimmed to separate the Au film from the Scotch tape. (f) The Scotch tape is removed, leaving the detached GNR/Au film on the target substrate. (g) The Au layer is etched away using a potassium iodide (KI) solution. (h) Only the GNRs remain on the substrate.

**XPS of Transferred 7-AGNRs Using LMPM-Assisted and Conventional Methods**

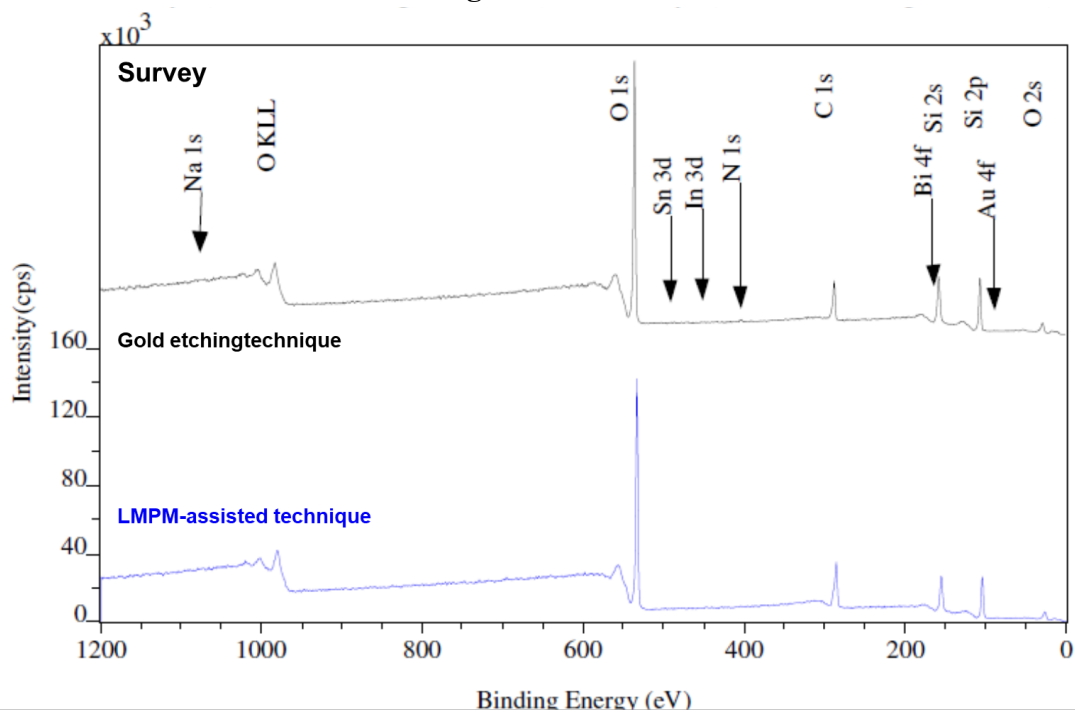

Figure S7: Comparison of XPS of transferred 7-AGNRs using LMPM-assisted and conventional methods.

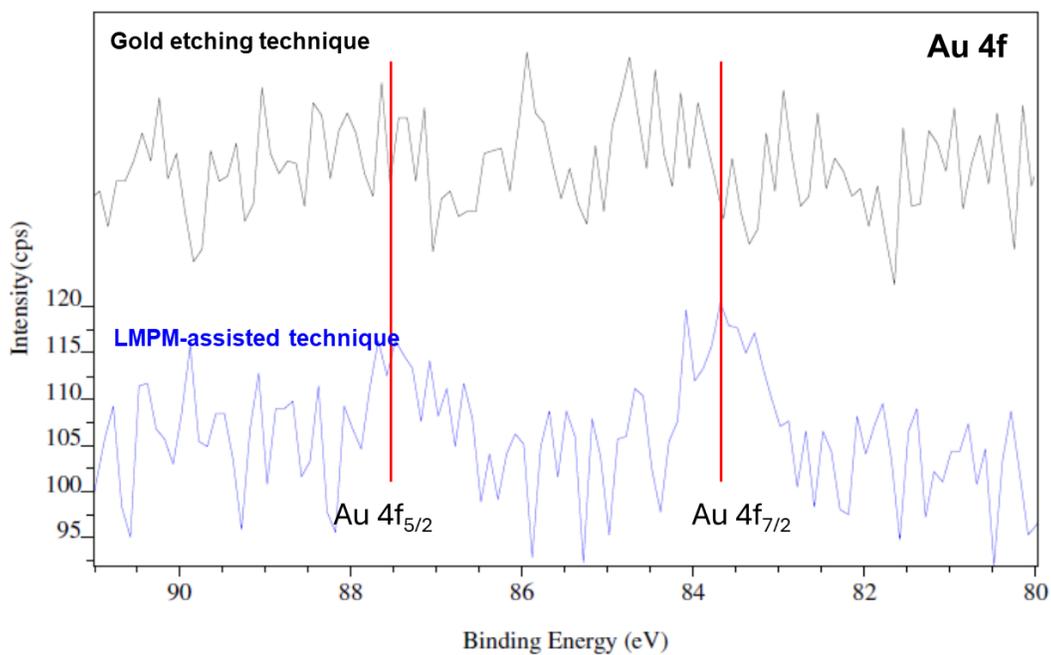

Figure S8: Expanded XPS spectra of the Au 4f region.

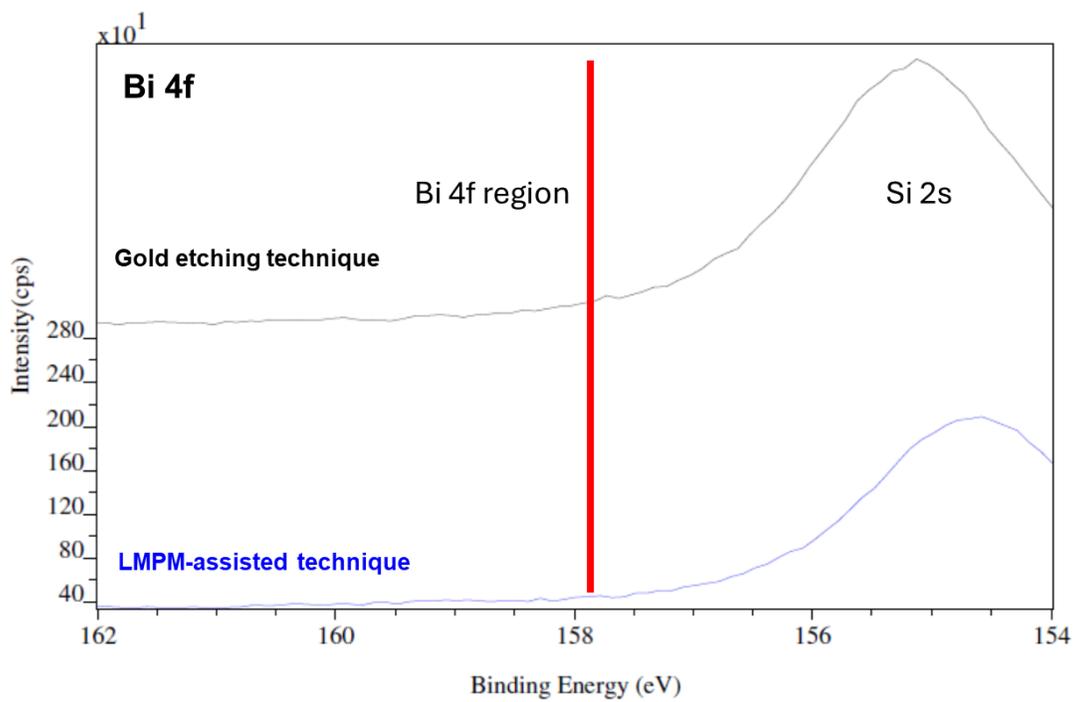

Figure S9: Expanded XPS spectra of the Bi 4f region.

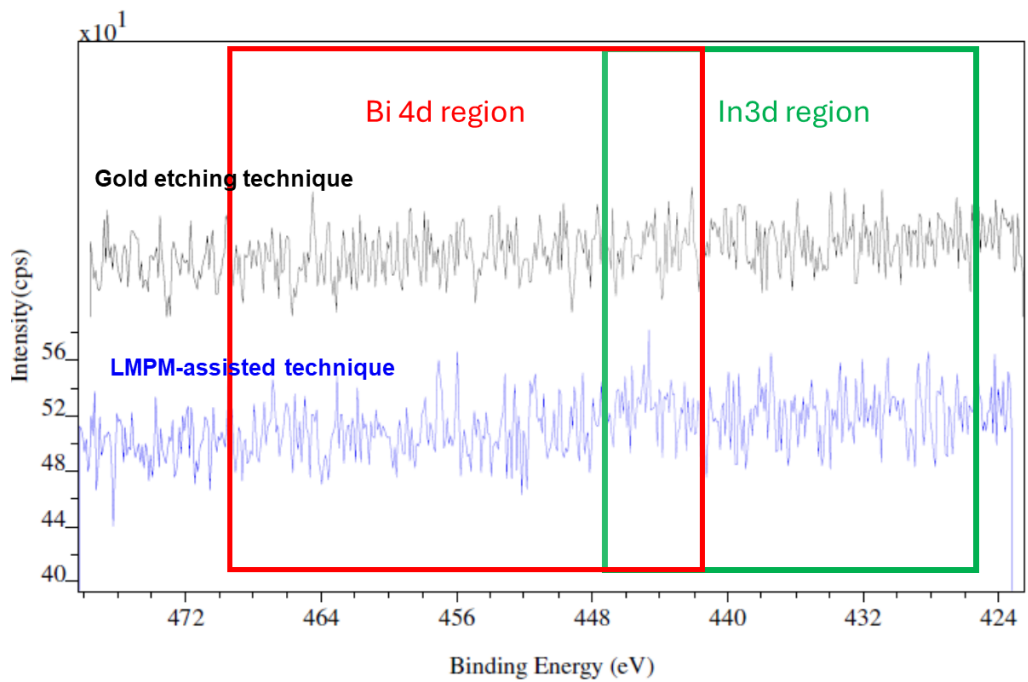

Figure S10: Expanded XPS spectra of the Bi 4d and In 3d region.

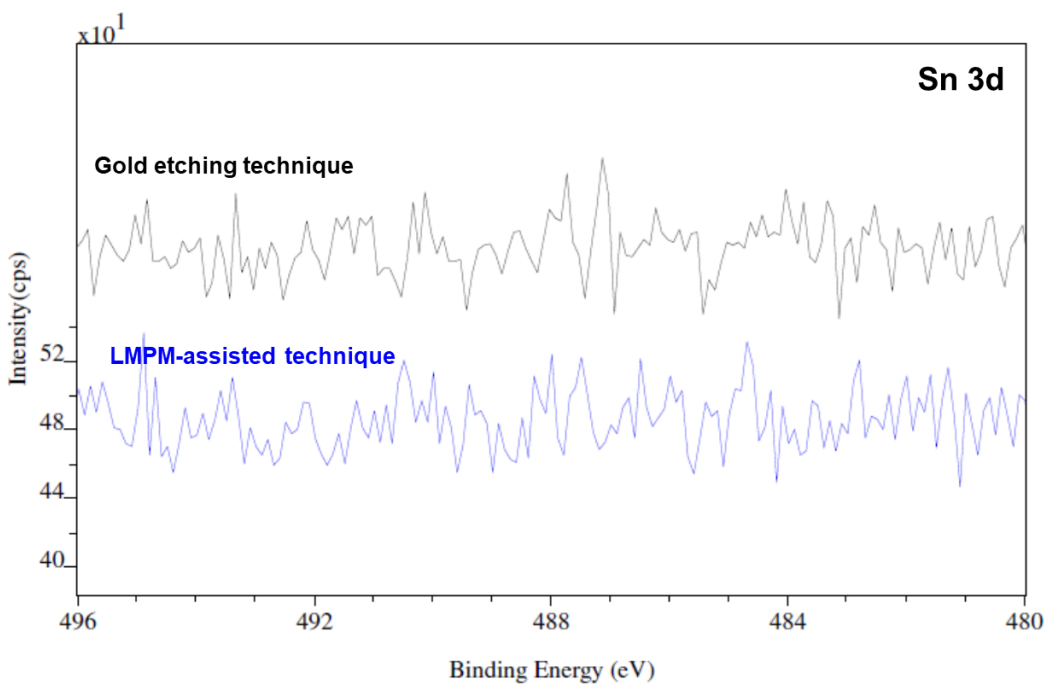

Figure S11: Expanded XPS spectra of the Sn 3d region.

## Characterization of Synthesized LaOCl Crystals

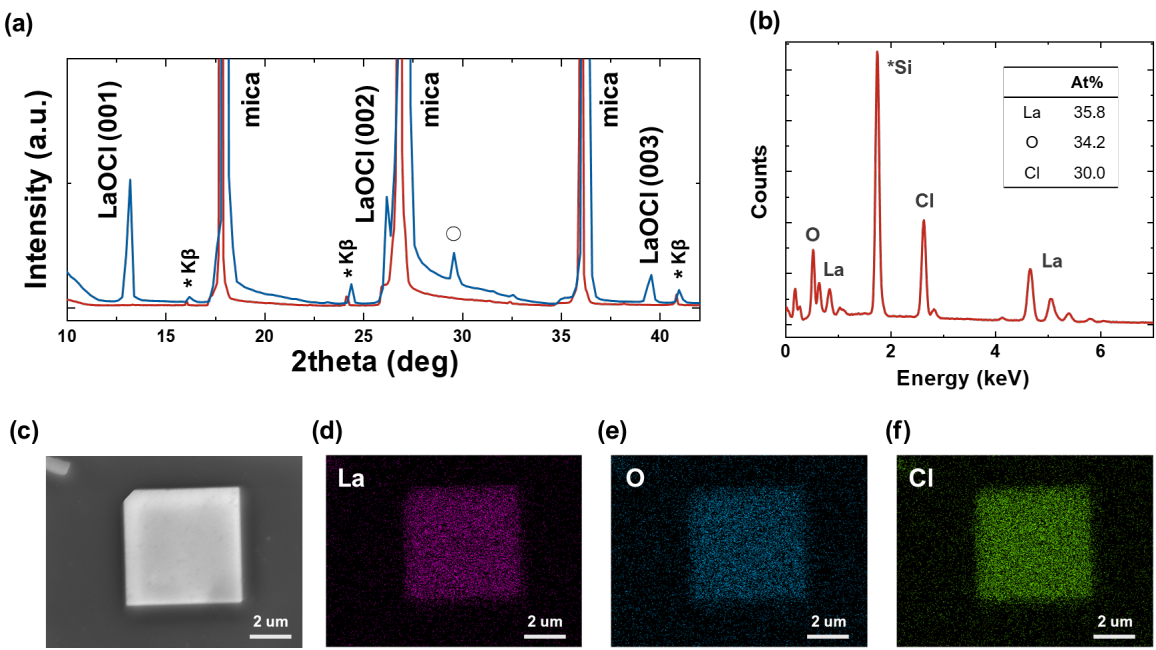

Figure S12: (a) XRD patterns of LaOCl deposited on mica substrate (blue) and bare mica substrate (red). (b) EDS spectrum of LaOCl on Si substrate. (c) SEM image of LaOCl sample with corresponding elemental EDS mapping of (d) La, (e) O, and (f) Cl.

## High Wettability of Field's Metal on PDMS

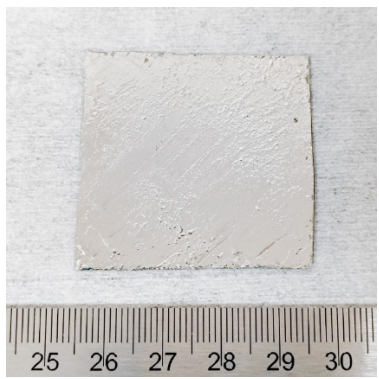

Figure S13: Optical image of LMPM on PDMS substrate. LMPM spreads thinly and uniformly over a large area compared to its behavior on glass surfaces as shown in Figure 2f.

## Raman Spectra of 9-AGNRs Transferred for FET Device Fabrication

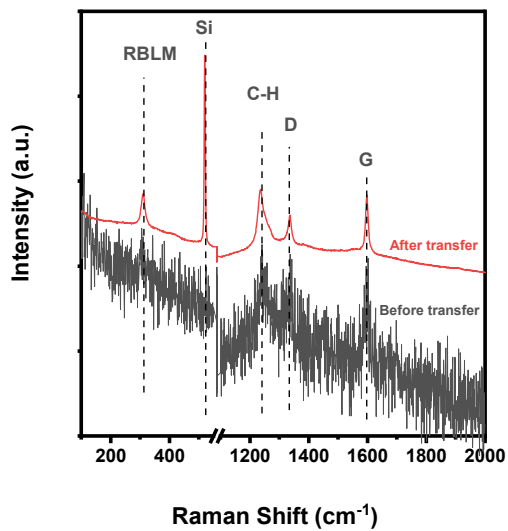

Figure S14: Raman spectra of graphene nanoribbons (9-AGNRs) before and after transfer from Au/mica substrate to pre-patterned chips.